\documentclass[amsmath,amssymb]{revtex4}
\usepackage{graphicx}
\usepackage{dcolumn}
\usepackage{bm}

\newcommand{\gsim}{\raisebox{0.2ex}{$\ > \kern -1.05em%
        \raisebox{-1.1ex}{$\sim$}\ $}}  
\begin{document}

\title{Effective potential analytic continuation calculations\\ 
of real time quantum correlation functions: 
Asymmetric systems
}

\author{Atsushi Horikoshi$^{1,2}$}
 \email{horikosi@cc.nara-wu.ac.jp}
\author{Kenichi Kinugawa$^{2}$}%
 \email{kinugawa@cc.nara-wu.ac.jp}
\affiliation{
$^{1}$Japan Science and Technology Agency 
\\and\\
$^{2}$Department of Chemistry, Faculty of Science, 
Nara Women's University,
 \\
Nara 630-8506, Japan}

\date{\today}

\begin{abstract}
We apply the effective potential analytic continuation (EPAC) method 
to one-dimensional asymmetric potential systems
to obtain 
the real time quantum correlation functions at various temperatures.
Comparing the EPAC results with the exact results,
we find that for an asymmetric anharmonic oscillator
the EPAC results are in very good agreement with the exact ones
at low temperature,
while this agreement becomes worse as the temperature increases.  
We also show that the EPAC calculation for a certain type of 
asymmetric potentials can be reduced to that for 
the corresponding symmetric potentials. 
\end{abstract}
\maketitle
\section{INTRODUCTION}
\hspace*{\parindent}
The imaginary time path integral formalism \cite{fh}
has been an important tool for studying  
static or dynamical properties of 
quantum statistical mechanical systems.
The static properties have successfully 
been calculated by means of the 
path integral Monte Carlo (PIMC) or 
path integral molecular dynamics (PIMD) technique \cite{bt,ce}.
However, these numerical methods have difficulties 
in computing the dynamical properties such as 
the real time quantum correlation functions,
because the analytic continuation from the imaginary time 
to the real time is a non-trivial 
procedure in the numerical calculations \cite{tb,si,gb}.

\par
As new quantum dynamics methods to calculate 
the real time quantum correlation functions 
at finite temperature,
the centroid molecular dynamics (CMD) method \cite{cv} and 
the effective potential analytic continuation (EPAC) method \cite{hk}
have recently been proposed. 
Both the methods use effective potentials, 
which can be numerically calculated
by means of the PIMC or PIMD technique.
In computing  
the real time quantum correlation functions,
the CMD method uses the {\it effective classical potential} \cite{fh,fk,gt},
while the EPAC method uses the {\it standard effective potential} 
which appears in the effective action formalism \cite{ep,riv}.
The effective potentials 
have been extensively used as powerful tools  
in the calculations of the static properties \cite{fk,gt,ep,riv,ctvv,co},
so that they are expected to be effective also
for the calculations of the dynamical properties.
In practice, the CMD method has been 
tested in low-dimensional systems \cite{cv,jv,kb,ra,reich}
and 
applied to  
various many-body molecular systems 
to yield many promising results \cite{appli,appli2}.
On the other hand, 
it has already been shown that 
the EPAC method works well
in a one-dimensional symmetric double well system \cite{hk}.
This success suggests that the EPAC should be very useful 
for the calculation of quantum dynamics of the systems 
in which quantum coherence is significant.
\par
Most of actual potentials appearing 
in chemical systems and condensed matter systems
have asymmetric shapes.
For example, asymmetric double well potentials
have been widely used to describe 
the proton transfer reactions \cite{proton}
or
the first-order phase transitions \cite{gold,asym}.
The other asymmetric potentials such as the Morse potential have also been
very important 
and have been repeatedly used in chemical physics
as a realistic model of the intramolecular vibration.
\par
In the present paper, the EPAC method is therefore applied 
to the evaluation of real time quantum correlation functions
for one-dimensional asymmetric systems.
We investigate whether the EPAC method can successfully reproduce 
the exact correlation functions for the Morse potential. 
We also show that the EPAC calculation for a certain type of 
asymmetric potentials can be reduced to that for 
the corresponding symmetric potentials. 
This property is called the {\it decoupling property}, 
which is quite useful for the discussion of the asymmetrization effect 
in the quantum statistical systems,
for instance,
the enhancement or suppression of the transition rate 
in chemical reaction \cite{proton}. 
Furthermore, the decoupling property
can aid us in reducing the computational cost 
required on the PIMD/PIMC calculation.
In this paper, we analytically prove that this property holds,
followed by the numerical examination of it.
\par
In Sec. II, we summarize the definition 
of a couple of types of effective potentials 
and review our EPAC method.
Moreover, we explicitly present the standard effective potential 
and the EPAC position autocorrelation function for a harmonic oscillator
as the simplest example. 
The EPAC analyses of one-dimensional systems with asymmetric potentials
are shown in Sec. III.
After showing the decoupling property of linear terms appearing 
in the classical potentials, 
we present the EPAC results for the  
asymmetric harmonic/anharmonic oscillators.
The conclusions are given in Sec. IV.
\par
\section{EFFECTIVE POTENTIAL ANALYTIC CONTINUATION METHOD}

\subsection{Effective potentials and analytic continuation}
\hspace*{\parindent}
We first introduce
the effective classical potential
defined by Feynman \cite{fh}. 
Consider a quantum system where a quantum particle of mass $m$ moves 
in a one-dimensional potential $V(q)$ at temperature $T$.
The quantum canonical partition function of this system 
is expressed
in terms of the imaginary time path integral
\begin{eqnarray}
{\cal Z}_{\beta}&=&\int^{\infty}_{-\infty}dq
\int^{q(\beta\hbar)=q}_{q(0)=q}{\cal D}q
~e^{-S_{E}/\hbar}
, \label{1}
\end{eqnarray}
where $\beta=1/k_{B}T$ and $S_{E}$ is the
Euclidean action functional
\begin{eqnarray}
S_{E}[q]=\int^{\beta\hbar}_{0}d\tau
\left[~\!\frac{1}{2}~\!m ~\!\dot{q}^2 + V(q)~\!\right]
. \label{2}
\end{eqnarray}
After a Fourier decomposition of the periodic paths
$q(\tau)=(1/\beta\hbar)\sum_{n=-\infty}^{\infty}
e^{-i\omega_{n}\tau}\tilde{q}(\omega_n)$ 
with the Matsubara frequencies 
$\omega_{n}=2\pi n/\beta\hbar$,
we get \cite{fk,kl}
\begin{eqnarray}
{\cal Z}_{\beta}&=&
\sqrt{\frac{m}{2\pi\beta\hbar^2}}~\!
\int^{\infty}_{-\infty}dq_{0}
\prod_{n=1}^{\infty}
\frac{m\omega_{n}^{2}}{\pi\beta\hbar^2}~\!
\int^{\infty}_{-\infty}\int^{\infty}_{-\infty}
d({\rm Re}\tilde{q}(\omega_n))d({\rm Im}\tilde{q}(\omega_n))
~e^{-S_{E}/\hbar}
, \label{3}
\end{eqnarray}
where $q_0$ is the path centroid (zero mode):
$q_{0}=\tilde{q}(\omega_{0})/\beta\hbar$.
Inserting an identity $1=\int^{\infty}_{-\infty}dq_{c}\delta(q_{0}-q_{c})$ 
into the integral in Eq. (\ref{3})
and integrating out all the Fourier modes, 
we obtain 
\begin{eqnarray}
{\cal Z}_{\beta}
&=&\sqrt{\frac{m}{2\pi\beta\hbar^2}}
\int^{\infty}_{-\infty}dq_{c}
~\!e^{-\beta V_{\beta}^{c}(q_{c})}
.\label{4}
\end{eqnarray}
Here $V_{\beta}^{c}(q_{c})$ is 
the effective classical potential \cite{fh,fk,gt},
which is a function of 
the position centroid variable $q_{c}$.
Among the integration of all the Fourier modes,
the zero mode integral $\int^{\infty}_{-\infty}dq_{0}$ yields 
only the replacement $q_{0}\to q_{c}$,  
so that the effective classical potential $V_{\beta}^{c}(q_{c})$
does not contain 
the fluctuation effect of the zero mode $q_{0}$.
Therefore the effective classical potential contains only the  
quantum fluctuation effects of $\tilde{q}(\omega_{n\ne 0})$ modes.
\par
Another type of effective potential \cite{ep,riv}
is briefly summarized next. 
At first, consider the quantum canonical partition function represented 
in terms of an imaginary time path integral  
in the presence 
of a constant external source $J$,
\begin{eqnarray}
{\cal Z}_{\beta}(J)&=&\int^{\infty}_{-\infty}dq
\int^{q(\beta\hbar)=q}_{q(0)=q}{\cal D}q
~e^{-S_{E}/\hbar}
~e^{\beta Jq_{0}}
=~e^{\beta w_{\beta}(J)},
\label{6}
\end{eqnarray}
where $w_{\beta}(J)$ is the generating function 
for the connected Green functions with zero energy.
From Eqs. (\ref{4}) and (\ref{6}),
the generating function is 
written in terms of $V_{\beta}^{c}(q_{c})$,  
\begin{eqnarray}
w_{\beta}(J)
&=& 
\frac{1}{\beta}\log\left[\sqrt{\frac{m}{2\pi\beta\hbar^{2}}}
\int^{\infty}_{-\infty}d q_{c}~\!
e^{-\beta V_{\beta}^{c}(q_{c})}~\!
e^{\beta J q_{c}}
\right]
.\label{7}
\end{eqnarray}
This can be rewritten in the phase-space centroid representation,
\begin{eqnarray}
w_{\beta}(J)
&=& 
\frac{1}{\beta}\log\left[
\int^{\infty}_{-\infty}\!
\int^{\infty}_{-\infty}~\!
\frac{d q_{c}d p_{c}}{2\pi\hbar}~\!
e^{\beta J q_{c}}~\!
e^{-\beta \{p_{c}^{2}/2m+V_{\beta}^{c}(q_{c})\}}
\right]
,\label{7a}
\end{eqnarray}
where $p_{c}$ is the momentum centroid variable. 
The ``conventional'' effective potential
$V_{\beta}(Q)$ is defined as the Legendre transform of 
the generating function $w_{\beta}(J)$ \cite{ep,riv},
\begin{eqnarray}
V_{\beta}(Q)&=&\sup_{J}~\!
\{~\!JQ-w_{\beta}(J)~\!\}.
\label{8}
\end{eqnarray}
We refer to this effective potential as the
standard effective potential. 
This is the leading order of the derivative expansion of 
the effective action, and contains
the effects of both quantum fluctuation and thermal fluctuation
in the quantum statistical system.
The {\it static} properties of the system can be reproduced 
from $V_{\beta}(Q)$ straightforwardly \cite{ep,riv}.
For example, 
one can evaluate the expectation value of the position operator $\hat{q}$ 
using the relation  
\begin{eqnarray}
\langle \hat{q}\rangle_{\beta}=Q_{\rm min},\label{10} 
\end{eqnarray}  
where $Q_{\rm min}$ is 
the position of the standard effective potential minimum.
The value of $Q_{\rm min}$ is determined by the stationary condition
$\left.{\partial V_{\beta}}/{\partial Q} 
\right|_{Q=Q_{\rm min}}=0$.
In the low temperature limit, 
the two effective potentials are identical \cite{fukuda,hs,owy},
while the ground state energy of the quantum system is given by
the minimum value of $V_{\beta}(Q)$ \cite{co,riv},
\begin{eqnarray}
E_{0}&=& \langle 0|\hat{H}|0\rangle
=\lim_{\beta\to \infty} V_{\beta}(Q_{\rm min})
.\label{12}
\end{eqnarray}
\par
On the other hand, 
one can also calculate 
the {\it dynamical} properties such as 
real time quantum correlation functions
from the standard effective potential $V_{\beta}(Q)$.
Using the effective action formalism \cite{ep,riv},
at first imaginary time quantum correlation functions 
are directly obtained 
from the standard effective potential $V_{\beta}(Q)$
with the leading order derivative expansion.
Then by means of the analytic continuation \cite{bell}
real time quantum correlation functions can  
be readily obtained from the imaginary time quantities. 
This is the EPAC \cite{hk}, 
an approximation method based on the effective action formalism.
Following the EPAC method, for example,  
the real time position autocorrelation function
$C_{\beta}(t)=\langle \hat{q}(t)\hat{q}(0)\rangle_{\beta}$
can be expressed as \cite{hk} 
\begin{eqnarray}
C_{\beta}(t)\simeq C^{\rm AC}_{\beta}(t)
&=&\left(\frac{\hbar}{2m\omega_{\beta}}
\coth\frac{\beta\hbar\omega_{\beta}}{2}\right)\cos\omega_{\beta}t
-i\left(~\!\frac{\hbar}{2m\omega_{\beta}}\right)\sin\omega_{\beta}t
+Q_{\rm min}^{2}
,\label{22}
\end{eqnarray}
where $\omega_{\beta}$ is the effective frequency, 
\begin{eqnarray}
\omega_{\beta}&=&\sqrt{\frac{1}{m}\left.\frac{\partial^2 V_{\beta}}
{\partial Q^2}\right|}_{Q=Q_{\rm min}}
.\label{18}
\end{eqnarray}
\subsection{Example: harmonic oscillator}
\hspace*{\parindent}
To illustrate the characteristics of the standard effective potential 
$V_{\beta}(Q)$ and the EPAC correlation function,
here we explicitly present the results for a simple system:
a quantum harmonic oscillator whose classical potential is given by
\begin{eqnarray}
V(q)=\frac{1}{2}~\!m\omega^{2}q^{2}
.\label{23}
\end{eqnarray}
In this system 
the effective classical potential can be written as \cite{kl}
\begin{eqnarray}
V_{\beta}^{c}(q_{c})
&=&\frac{1}{2}m\omega^{2}q_{c}^{2}
+\frac{1}{\beta}\log
\left(\frac{\sinh(\beta\hbar\omega/2)}{\beta\hbar\omega/2}
\right)
.\label{25}
\end{eqnarray}
Then we can evaluate 
the generating function $w_{\beta}(J)$ from 
Eqs. (\ref{7}) and (\ref{25}),
\begin{eqnarray}
w_{\beta}(J)
&=& 
\frac{1}{2m\omega^{2}}J^{2}-
\frac{1}{\beta}\log
\left(2\sinh\frac{\beta\hbar\omega}{2}\right),
\label{26}
\end{eqnarray}
to obtain eventually the standard effective potential 
as the Legendre transform of $w_{\beta}(J)$ 
\begin{eqnarray}
V_{\beta}(Q)
&=&\frac{1}{2}m\omega^{2}Q^{2}+
\frac{1}{\beta}\log
\left(2\sinh\frac{\beta\hbar\omega}{2}\right).
\label{27}
\end{eqnarray}
The minimum of $V_{\beta}(Q)$ is located
at the point $Q=Q_{\rm min}=0$,
and the effective frequency is temperature-independent: 
$\omega_{\beta}=\omega$. 
This potential minimum $V_{\beta}(Q=0)$ is equal to 
the free energy of the quantum harmonic oscillator.   
\par

\begin{figure}
\includegraphics[width=80mm,height=80mm]{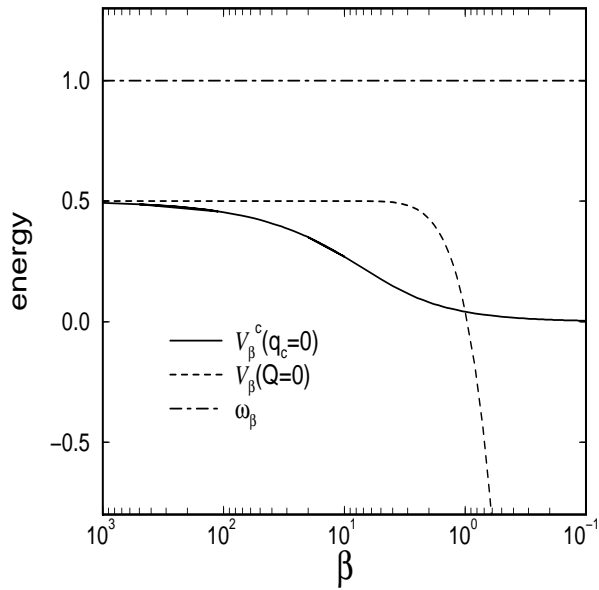}
\vspace{0mm}
\caption{
The inverse temperature $\beta$ dependence 
of  $V_{\beta}^{c}(q_{c}=0)$, $V_{\beta}(Q=0)$, 
and $\omega_{\beta}$ for the quantum harmonic oscillator (\ref{23})
with the parameters $\hbar=k_{B}=m=\omega=1$.
Note that $\omega_{\beta}$ of the harmonic oscillator is
independent of the temperature. 
}
\label{fig:harm}
\end{figure}

\par
In fact, in the low temperature limit, 
the potential minimum $V_{\beta}(Q=0)$ is equal to 
that of the effective classical potential $V_{\beta}^{c}(q_{c}=0)$,
and they give 
the ground state energy of the quantum harmonic oscillator,
\begin{eqnarray}
\lim_{\beta\to \infty} V_{\beta}(Q=0)&=& 
\lim_{\beta\to \infty} V_{\beta}^{c}(q_{c}=0)=\frac{\hbar\omega}{2}
.\label{28}
\end{eqnarray}
On the other hand, in the high temperature limit ($\beta\to 0$),
these effective potentials exhibit quite different features.
The standard effective potential minimum $V_{\beta}(Q=0)$
diverges to negative infinity, while 
the effective classical potential minimum 
$V_{\beta}^{c}(q_{c}=0)$
converges to zero.  
This is because 
the standard effective potential $V_{\beta}(Q)$
includes the effects of 
both quantum fluctuation and thermal fluctuation,
while the effective classical potential $V_{\beta}^{c}(q_{c})$
includes the effect of quantum fluctuation only. 
In Fig. \ref{fig:harm} we show 
the inverse temperature $\beta$ dependence of 
$V_{\beta}^{c}(q_{c}=0)$, $V_{\beta}(Q=0)$, and $\omega_{\beta}$
with the parameters $\hbar=k_{B}=m=\omega=1$.
\par
Finally we obtain the EPAC position autocorrelation function 
for the quantum harmonic oscillator
\begin{eqnarray}
C^{\rm AC}_{\beta}(t)
&=&\left(\frac{\hbar}{2m\omega}
\coth\frac{\beta\hbar\omega}{2}\right)\cos\omega t
-i\left(~\!\frac{\hbar}{2m\omega}\right)\sin\omega t
.\label{29}
\end{eqnarray}
This is equal to the exact correlation function. 
The EPAC method is thus exact for harmonic systems
as well as the CMD \cite{comment0}.

\section{APPLICATION TO ASYMMETRIC SYSTEMS}
\hspace*{\parindent}
In our previous paper \cite{hk}, the usefulness of the EPAC method
has been shown for a $Z_{2} (q\leftrightarrow-q)$ symmetric 
double well potential system in comparison with the CMD result.
We expect that the EPAC method may also be useful for 
general anharmonic quantum systems.
In the present section, we embark on a project to
investigate how the EPAC method 
works in a $Z_{2}$ asymmetric system 
with the classical potential \cite{cv,jv},
\begin{eqnarray}
V(q)=\frac{1}{2}~\!q^{2}+
\frac{1}{10}~\!q^{3}+
\frac{1}{100}~\!q^{4}
,\label{30}
\end{eqnarray}
where natural units $\hbar=k_{B}=m=1$ are employed.
This potential is given by a polynomial approximation
to a Morse potential, 
\begin{eqnarray}
V(q)=\frac{25}{2}~\!\left(
1-e^{(1/5)q}
\right)^{2}
\label{31}
\end{eqnarray} 
regarding $q$ as the small deviation 
from the minimum $q=0$ \cite{comment1}.
This polynomial approximation is very good for $\beta\gsim 10$.  
The Morse potential (\ref{31}) corresponds to, for example, 
the parameters denoting 
the intramolecular vibration of the HCl molecule \cite{herz}.
For this system the inverse temperature $\beta=10$
corresponds to the temperature $T\sim 400 ~{\rm K}$.
\par
On the other hand, by a variable shift
$x=q+\frac{5}{2}$,
we can transform the potential (\ref{30}) to the form
\begin{eqnarray}
\bar{V}(x)=\frac{125}{64}
-\frac{5}{4}~\!x
+\frac{1}{8}~\!x^{2}
+\frac{1}{100}~\!x^{4}
,\label{33}
\end{eqnarray} 
which is a potential asymmetrized by 
the linear term $-\frac{5}{4}~\!x$ only.
Such additional linear terms exhibit interesting feature
in the effective potential analyses;
they are irrelevant both 
to the path integral calculation (\ref{3})
and to the Legendre transformation (\ref{8}).
This property ensures that 
the standard effective potential for the asymmetric potential (\ref{33})
can readily be given by that for the corresponding symmetric potential
without the linear term.
This is the decoupling property, which has been discussed 
for other types of effective potentials
in the context of 
the nonperturbative renormalization group \cite{asym,nprg}.
In the following we give a proof of this property
for the standard effective potential, 
and then apply it to the evaluation of 
the standard effective potentials for 
an asymmetric harmonic oscillator and 
the asymmetric anharmonic oscillator (\ref{33}).

\subsection{Decoupling property of the additional linear term}
\hspace*{\parindent}
Consider a quantum system described 
by a classical potential of the form
\begin{eqnarray}
V(q)=U(q)+fq
,\label{34}
\end{eqnarray} 
where $U(q)$ is any classical potential and $f$ is a constant. 
The effective classical potential can be written as
\begin{eqnarray}
V_{\beta}^{c}(q_{c})=U_{\beta}^{c}(q_{c})+fq_{c}
,\label{35}
\end{eqnarray}  
where $U_{\beta}^{c}(q_{c})$ is the effective classical potential 
for $U(q)$. 
This is because in the Euclidean action $S_{\rm E}$
the linear term 
consists of only the zero mode $q_{0}$,
\begin{eqnarray}
\int^{\beta\hbar}_{0}d\tau~\!fq(\tau)=f\beta\hbar q_{0}
,\label{36}
\end{eqnarray} 
and never generate or receive any quantum correction
when we integrate out all the Fourier modes to evaluate 
the effective classical potential.
\par
Next, from Eqs. (\ref{7}) and (\ref{35})
the generating function for $V(q)$ is given as 
\begin{eqnarray}
w_{\beta}(J)&=&\frac{1}{\beta}\log\left[
\sqrt{\frac{m}{2\pi\beta\hbar^2}}
\int^{\infty}_{-\infty}d q_{c}~\! 
e^{-\beta~\!U_{\beta}^{c}(q_{c})}
e^{\beta~\!(J-f) q_{c}}\right]
.\label{37}
\end{eqnarray}
Supposing that $w_{\beta}^{U}$ is 
the generating function for $U(q)$,
an evident relation 
$w_{\beta}(J)=w_{\beta}^{U}(J-f)$ holds.
Furthermore, supposing $J^{*}(Q)$ is the solution of the equation
$Q=\partial w_{\beta}/\partial J$ 
and $J_{U}^{*}(Q)$ is that of $Q=\partial w^{U}_{\beta}/\partial J$,
we get a relation 
$J^{*}(Q)=J_{U}^{*}(Q)+f$.
The standard effective potential for $V(q)$
can therefore be written as 
\begin{eqnarray}
V_{\beta}(Q)
&=&\{~\!JQ-w_{\beta}(J)~\!\}|_{J=J^{*}(Q)}\nonumber\\
&=&J^{*}_{U}(Q)Q+fQ
-\frac{1}{\beta}\log\left[
\sqrt{\frac{m}{2\pi\beta\hbar^{2}}}
\int^{\infty}_{-\infty}d q_{c}~\!
e^{-\beta U_{\beta}^{c}(q_{c})}
e^{\beta J^{*}_{U}(Q)q_{c}}
\right]
\nonumber\\
&=&
U_{\beta}(Q)+fQ
,\label{40}
\end{eqnarray}
where $U_{\beta}(Q)$ is the standard effective potential for $U(q)$.
Equations (\ref{35}) and (\ref{40}) ensure that
the functional form of the linear term
remains unchanged 
through the path integration and the Legendre transformation,
and that it never generates any quantum or thermal correction to $U(q)$.
That is, the linear term in the classical potential 
decouples from the effective potential calculations.
\par
Using Eq. (\ref{40}),
we can directly obtain the standard effective potential $V_{\beta}(Q)$ 
from $U_{\beta}(Q)$.
This means that
we have only to carry out 
the path integration and the Legendre transformation 
for $U(q)$.
This decoupling property is especially useful 
when $U(q)$ is a $Z_{2} (q\leftrightarrow -q)$ symmetric potential,
because we can simply discuss the asymmetrization effect 
as the additional linear term effect on $U_{\beta}(Q)$. 
Furthermore, 
the $Z_{2}$ symmetry reduces 
the computational cost required on the PIMD/PIMC calculation 
for the $Z_{2}$ symmetric $U(q)$. 
This is because the required cite number in $q_{c}$ space 
is reduced by half by the $Z_{2}$ symmetry \cite{owy}.    

\subsection{Asymmetric harmonic oscillator}
\hspace*{\parindent}
As a simple example of the system with the 
additional linear term (\ref{34}), 
we consider an asymmetric harmonic oscillator 
\begin{eqnarray}
V(q)=\frac{1}{2}~\!m\omega^{2}q^{2}+fq
.\label{41}
\end{eqnarray}
From Eqs. (\ref{27}) and (\ref{40})
we readily obtain 
the standard effective potential 
\begin{eqnarray}
V_{\beta}(Q)
&=&\frac{1}{2}m\omega^{2}Q^{2}+fQ
+\frac{1}{\beta}\log
\left(2\sinh\frac{\beta\hbar\omega}{2}\right)
.\label{43}
\end{eqnarray}
The minimum of $V_{\beta}(Q)$ is at the point 
$Q=Q_{\rm min}=-f/m\omega^{2}$,
and the effective frequency 
is still temperature-independent: 
$\omega_{\beta}=\omega$. 
The value $Q_{\rm min}$ coincides with the exact position expectation value 
$\langle \hat{q}\rangle_{\beta}$,
while the minimum value, 
\begin{eqnarray}
V_{\beta}(Q_{\rm min})
=-\frac{f^{2}}{2m\omega^{2}}+\frac{1}{\beta}\log
\left(2\sinh\frac{\beta\hbar\omega}{2}\right),
\label{45}
\end{eqnarray}
gives the exact ground state energy in the low temperature limit,
\begin{eqnarray}
E_{0}=\lim_{\beta\to\infty}V_{\beta}(Q_{\rm min})
=-\frac{f^{2}}{2m\omega^{2}}+\frac{\hbar\omega}{2}.
\label{46}
\end{eqnarray}
\par
The EPAC position autocorrelation function is then obtained as 
\begin{eqnarray}
C^{\rm AC}_{\beta}(t)
&=&\left(\frac{\hbar}{2m\omega}
\coth\frac{\beta\hbar\omega}{2}\right)\cos\omega t
-i\left(~\!\frac{\hbar}{2m\omega}\right)\sin\omega t
+\frac{f^{2}}{m^{2}\omega^{4}}
.\label{47}
\end{eqnarray}
This present result is also equal to 
the exact correlation function.

\subsection{Asymmetric anharmonic oscillator}
\hspace*{\parindent}
In this section we attempt to apply  
the EPAC method to 
the asymmetric anharmonic system with the potential (\ref{30}).
To obtain the parameters $Q_{\rm min}$ and $\omega_{\beta}$
needed for the EPAC correlation function (\ref{22}),
now we propose the following computational schemes. 
\par\vspace{10mm}
{\it Scheme (A)}.  By carrying out the path integration and 
the Legendre transformation 
for the classical asymmetric potential (\ref{30}),
we obtain the standard effective potential $V_{\beta}(Q)$.
We then fix the parameters $Q_{\rm min}$ and $\omega_{\beta}$
by minimizing $V_{\beta}(Q)$. 
\par
{\it Scheme (B)}. We first carry out the path integration and the Legendre 
transformation for the symmetric part in the 
converted potential (\ref{33}), 
\begin{eqnarray}
\bar{U}(x)=\frac{125}{64}
+\frac{1}{8}~\!x^{2}
+\frac{1}{100}~\!x^{4}
,\label{48}
\end{eqnarray} 
to obtain the symmetric standard effective potential $\bar{U}_{\beta}(X)$.
Using Eq. (\ref{40}) with $f=-\frac{5}{4}$, 
we can write the asymmetric standard effective potential as 
\begin{eqnarray}
\bar{V}_{\beta}(X)
&=&
\bar{U}_{\beta}(X)-\frac{5}{4}X
.\label{49}
\end{eqnarray}
We then obtain the minimum point $X_{\rm min}$
and the effective frequency $\bar{\omega}_{\beta}$ 
by minimizing $\bar{V}_{\beta}(X)$. 
The parameters $Q_{\rm min}$ and $\omega_{\beta}$
are obtained from the relations 
$Q_{\rm min}=X_{\rm min}-\frac{5}{2}$
and 
$\omega_{\beta}=\bar{\omega}_{\beta}$.
\par\vspace{10mm}
In the following 
we perform the numerical calculation with scheme (A) 
to test the EPAC method itself. 
The numerical calculation with scheme (B) is also performed
to check the decoupling property (\ref{40}). 
\par
\begin{figure}
\begin{tabular}{c}
\includegraphics[width=80mm]{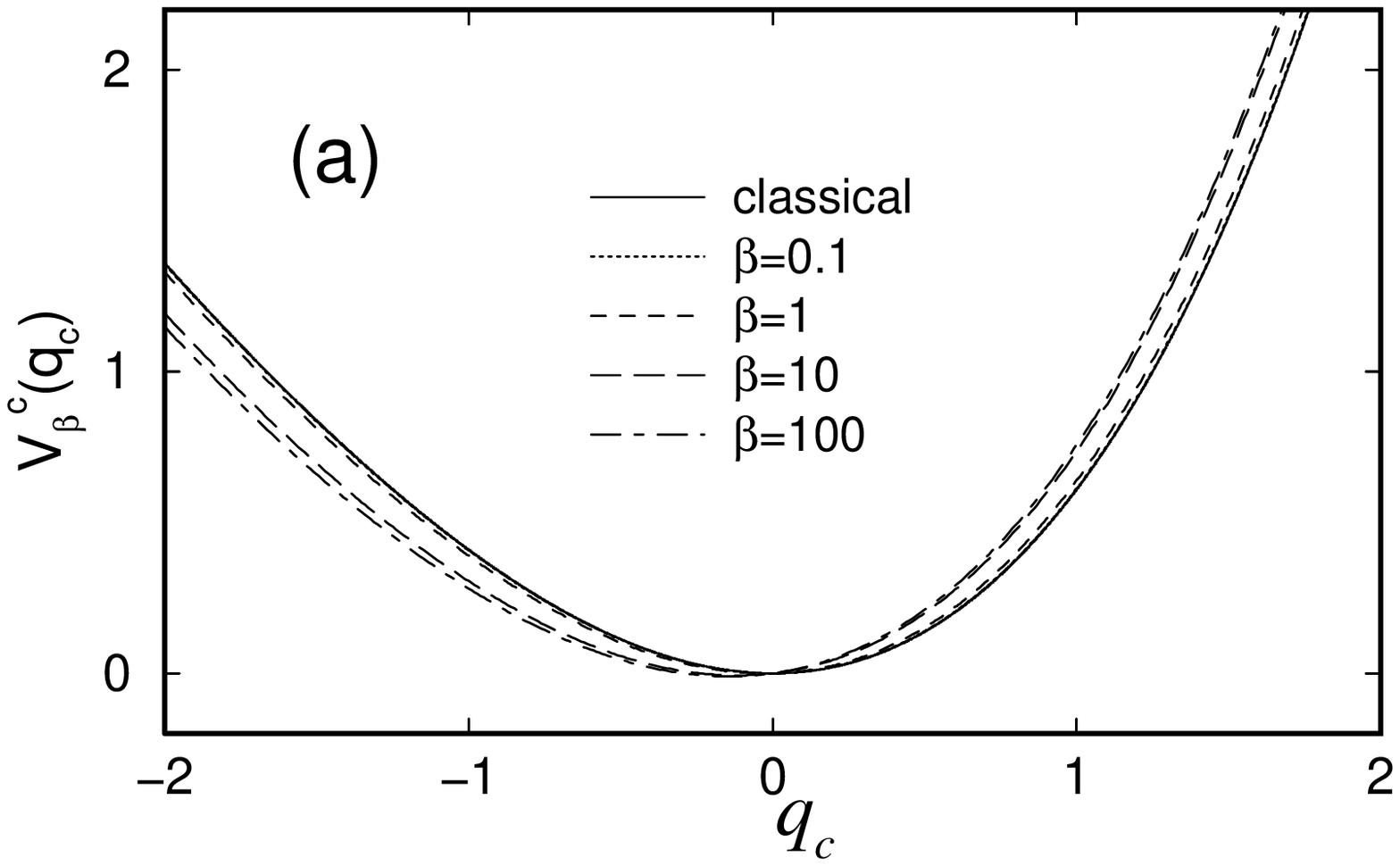}\\
\includegraphics[width=80mm]{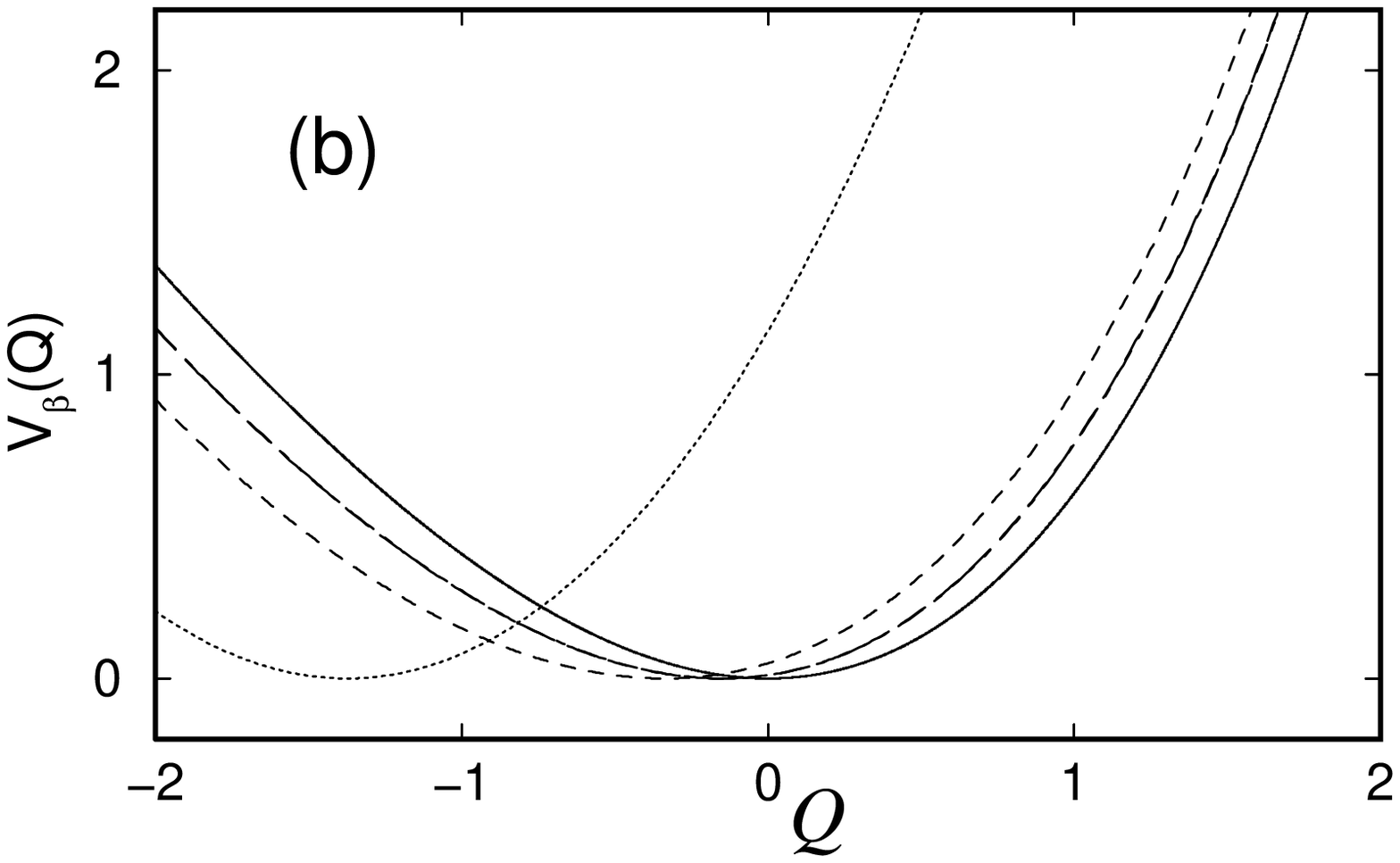}\\
\end{tabular}
\vspace{0mm}
\caption{
The inverse temperature $\beta$ dependence of
the effective potentials
for the asymmetric classical potential (\ref{30}):
(a) the effective classical potential $V_{\beta}^{c}(q_{c})$
and 
(b) the standard effective potential $V_{\beta}(Q)$. 
In this plot 
we set $V_{\beta}^{c}(0)=0$ and $V_{\beta}(Q_{\rm min})=0$.
}
\label{fig:epv}
\end{figure}
\par
\begin{figure}
\begin{tabular}{c}
\includegraphics[width=80mm]{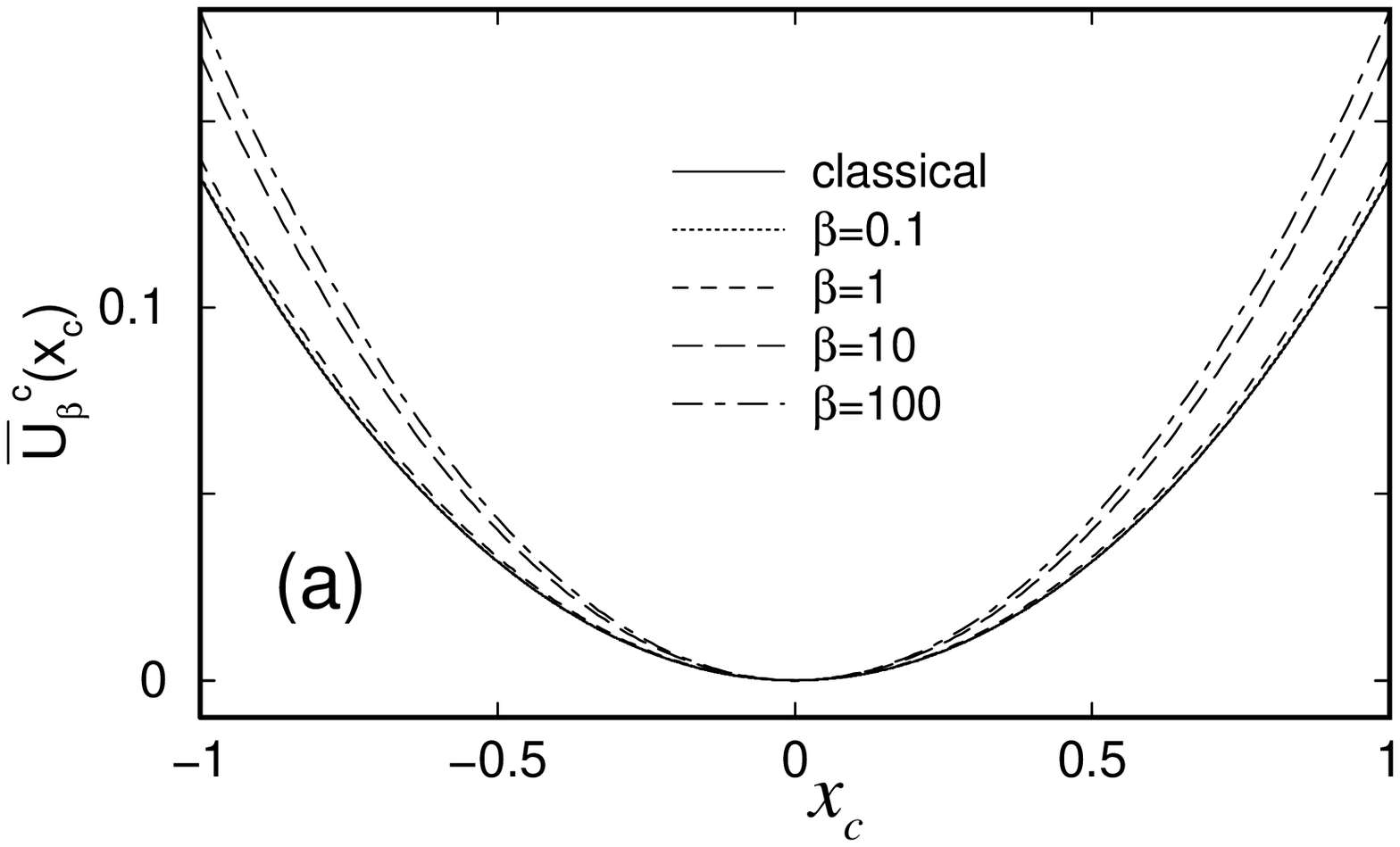}\\
\includegraphics[width=80mm]{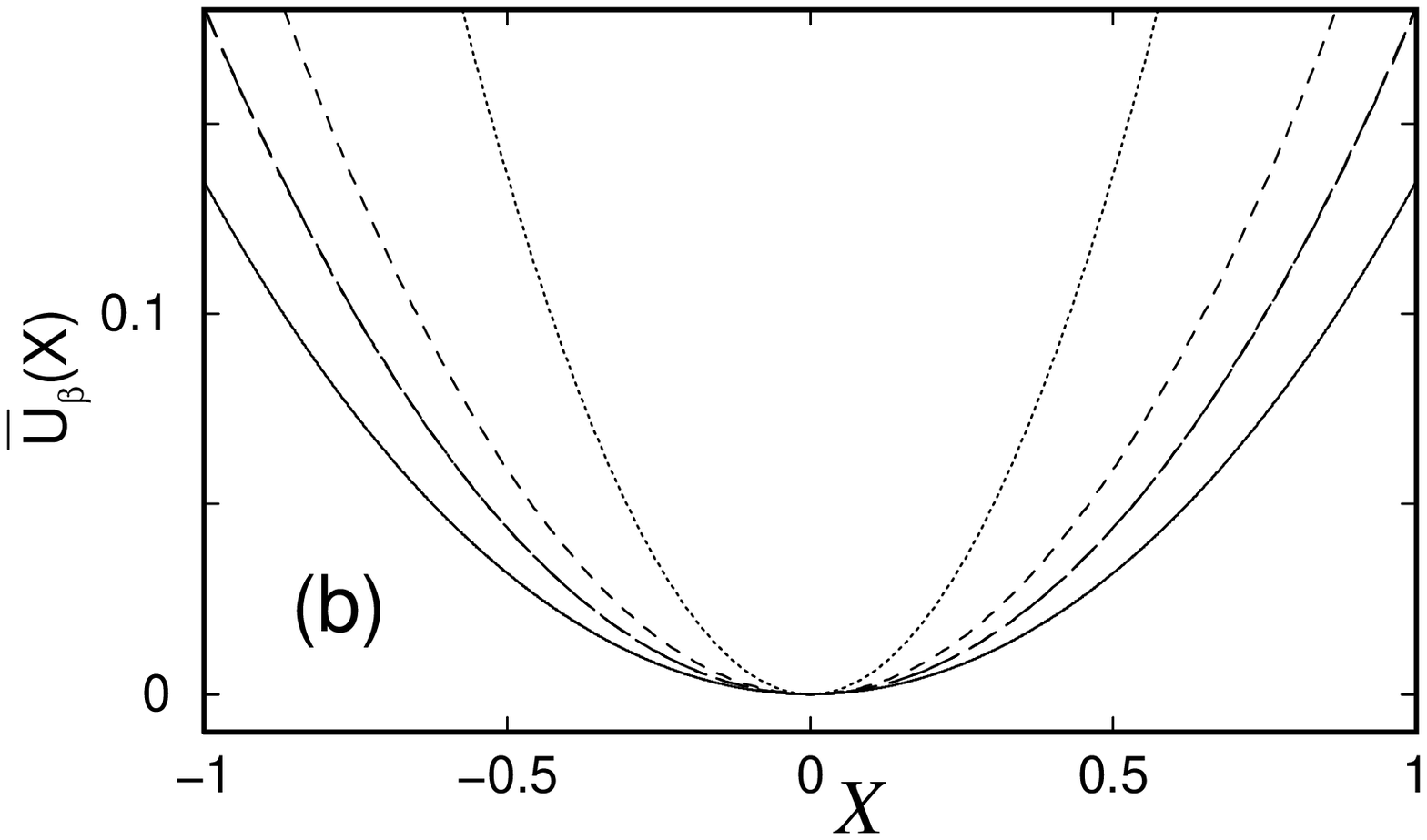}\\
\end{tabular}
\vspace{0mm}
\caption{
The inverse temperature $\beta$ dependence of
the effective potentials
for the symmetric classical potential (\ref{48}):
(a) the effective classical potential $\bar{U}_{\beta}^{c}(x_{c})$
and 
(b) the standard effective potential $\bar{U}_{\beta}(X)$.
In this plot
we set $\bar{U}(0)=0$,
$\bar{U}_{\beta}^{c}(0)=0$, and $\bar{U}_{\beta}(X_{\rm min})=0$.
}
\label{fig:epu}
\end{figure}
\par

\begin{figure}
\includegraphics[width=80mm,height=80mm]{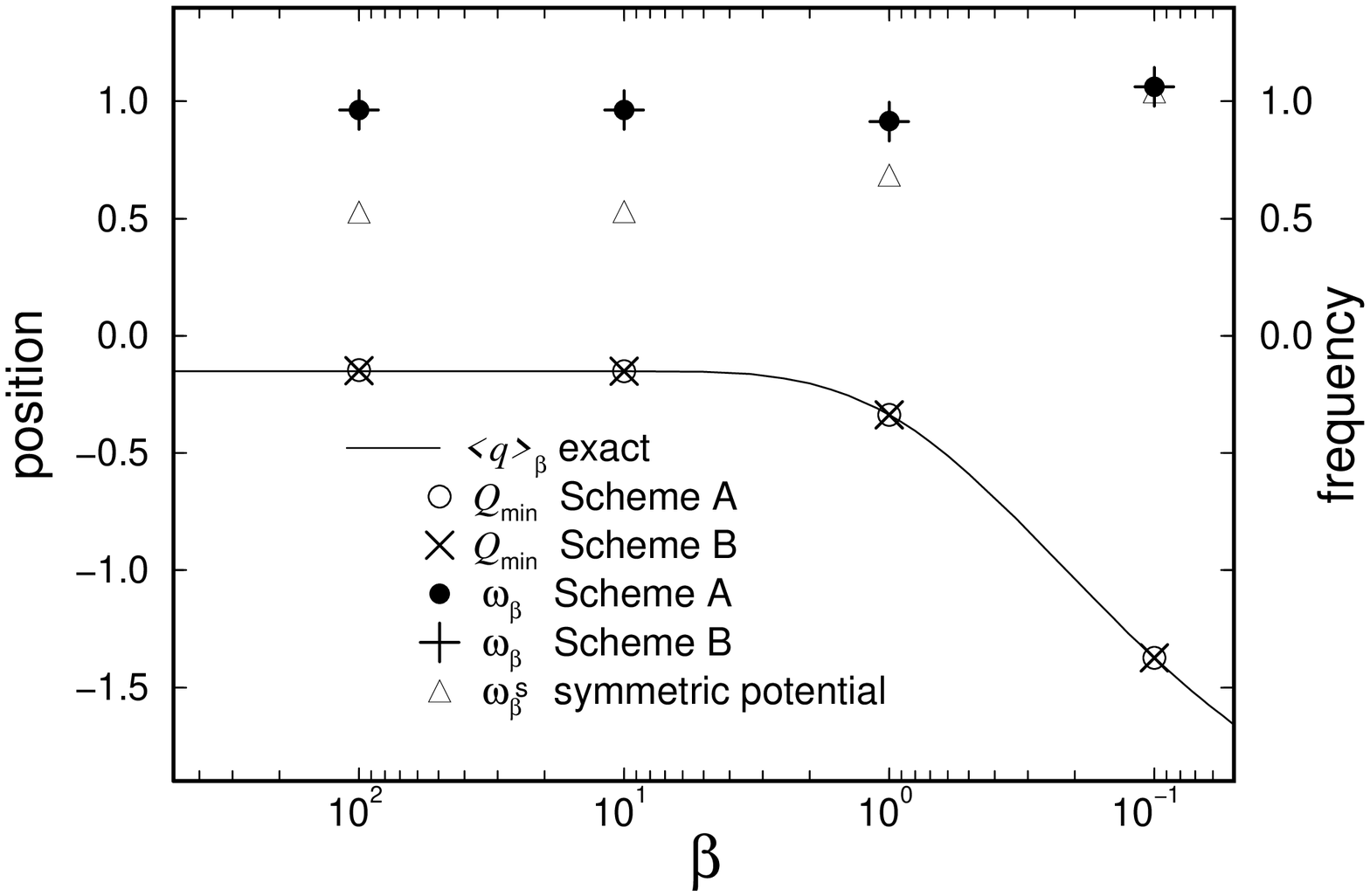}
\vspace{0mm}
\caption{
The inverse temperature $\beta$ dependence of 
the minimum point $Q_{\rm min}$ and
the effective frequency $\omega_{\beta}$.
The exact position expectation value 
$\langle \hat{q}\rangle_{\beta}$ and
the effective frequency of 
the symmetric effective potential $\omega_{\beta}^{s}$
are also plotted.
}
\label{fig:qw}
\end{figure}
\par
To evaluate the standard effective potentials 
$V_{\beta}(Q)$ in scheme (A) and 
$\bar{U}_{\beta}(X)$ in scheme (B)
using Eqs. (\ref{7}) and (\ref{8}),
we need to precompute the effective classical potentials
$V_{\beta}^{c}(q_{c})$ and $\bar{U}_{\beta}^{c}(x_{c})$,
respectively \cite{comment2}.
Employing the PIMD technique and the fitting procedure 
described in Ref. \onlinecite{hk},
we calculated $V_{\beta}^{c}(q_{c})$ and 
$\bar{U}_{\beta}^{c}(x_{c})$ 
at the inverse temperature $\beta=0.1,1,10$, and $100$.
We then calculated the standard effective potential
$V_{\beta}(Q)$ and $\bar{U}_{\beta}(X)$ by carrying out 
the numerical integration (\ref{7})
and
the numerical Legendre transformation (\ref{8}).
As for scheme (A),
Fig. \ref{fig:epv} shows the evaluated effective potentials
for the asymmetric potential $V(q)$ [Eq. (\ref{30})].
As seen in Fig. \ref{fig:epv}(a) and \ref{fig:epv}(b), 
at the low temperature $\beta=100$,
the standard effective potential $V_{\beta}(Q)$
has almost the same shape as 
the effective classical potential $V_{\beta}^{c}(q_{c})$. 
This is a general property of  
the effective potentials \cite{fukuda,hs,owy}.
With regard to Scheme (B),
the evaluated effective potentials for 
the symmetric part $\bar{U}(x)$ [Eq. (\ref{48})]
are shown in Fig. \ref{fig:epu}.
It is also seen in this figure that   
$\bar{U}_{\beta}^{c}(x_{c})$ and $\bar{U}_{\beta}(X)$
have more similar shape at lower temperature,
as in Fig. \ref{fig:epv}.
\par
Next we evaluated the parameters
$Q_{\rm min}$ and $\omega_{\beta}$.
With scheme (A) we directly obtained them 
from the computed $V_{\beta}(Q)$,
while in scheme (B) 
they were obtained from  
the evaluated effective potential $\bar{U}_{\beta}(X)$ 
using the relations 
$Q_{\rm min}=X_{\rm min}-\frac{5}{2}$
and 
$\omega_{\beta}=\bar{\omega}_{\beta}$.
Figure \ref{fig:qw} shows 
$Q_{\rm min}$ and $\omega_{\beta}$ 
calculated by means of scheme (A) and (B),
and also shows the exact result 
\begin{eqnarray}
\langle\hat{q}\rangle_{\beta}=\frac{1}{{\cal Z}_{\beta}}\sum_{n}
e^{-\beta E_{n}}
\langle n|\hat{q}|n\rangle
,\label{53}
\end{eqnarray}
where the energy eigenstates $|n\rangle$ and 
the energy eigenvalues $E_{n}$ were obtained   
by a numerical integration of the Schr\"odinger equation. 
In this figure,
we also plotted the effective frequency of 
the symmetric effective potential $\bar{U}_{\beta}(X)$:
$\omega_{\beta}^{s}=
\sqrt{\partial^{2} \bar{U}_{\beta}(X)/\partial X^{2} |}_{X=0}$.
It is found that
at each temperature the quantities computed in scheme (B) 
are in very good agreement with the corresponding ones in scheme (A).
This indicates that the relation (\ref{40}) actually holds 
in the numerical PIMD calculation and 
the numerical Legendre transformation. 
We can also see that each $Q_{\rm min}$  
agrees very well 
with the corresponding exact value $\langle\hat{q}\rangle_{\beta}$;
the effective potential approach to such a static property 
works successfully   
in this asymmetric system.
\par
Here we give a comment on the asymmetrization effect.
As seen in Fig. \ref{fig:qw},
each effective frequency $\omega_{\beta}$($=\bar{\omega}_{\beta}$) 
is larger than 
$\omega_{\beta}^{s}$ at each temperature,
that is, 
the asymmetrization effect enhances 
the effective frequency of the system.
This property can be simply understood as follows:
For the weakly anharmonic system, 
the symmetric standard effective potential $\bar{U}_{\beta}(X)$
can be well approximated by a power series expansion 
to the fourth order,
$\bar{U}_{\beta}(X)=
C+\frac{1}{2}\omega_{\beta}^{s2}X^2+\frac{1}{4}\lambda_{\beta}X^4$
with a constant $C$ and a positive coupling constant $\lambda_{\beta}$.
Then the asymmetric standard effective potential 
can be written as $\bar{V}_{\beta}(X)=C-\frac{5}{4}X+
\frac{1}{2}\omega_{\beta}^{s2}X^2+\frac{1}{4}\lambda_{\beta}X^4$.
Supposing $X_{\rm min}$ is the minimum point of $\bar{V}_{\beta}(X)$,
we obtain the effective frequency of $\bar{V}_{\beta}(X)$ as
$\bar{\omega}_{\beta}=
\sqrt{\omega_{\beta}^{s2}+3\lambda_{\beta}X_{\rm min}^{2}}$. 
Since the parameters $\omega_{\beta}^{s2}$ and 
$\lambda_{\beta}$ are positive,
we get an inequality $\bar{\omega}_{\beta}\geq\omega_{\beta}^{s}$, 
where the equality holds in the harmonic limit $\lambda_{\beta}\to 0$. 
Figure \ref{fig:qw} also shows that 
$\omega_{\beta}$($=\bar{\omega}_{\beta}$)
becomes closer to $\omega_{\beta}^{s}$
as the temperature increases.
This implies that the standard effective potentials $\bar{U}_{\beta}(X)$,
$\bar{V}_{\beta}(X)$, and $V_{\beta}(Q)$ become  
effectively harmonic ones at the higher temperature.
\par
\begin{figure}
\begin{tabular}{c}
\includegraphics[width=80mm]{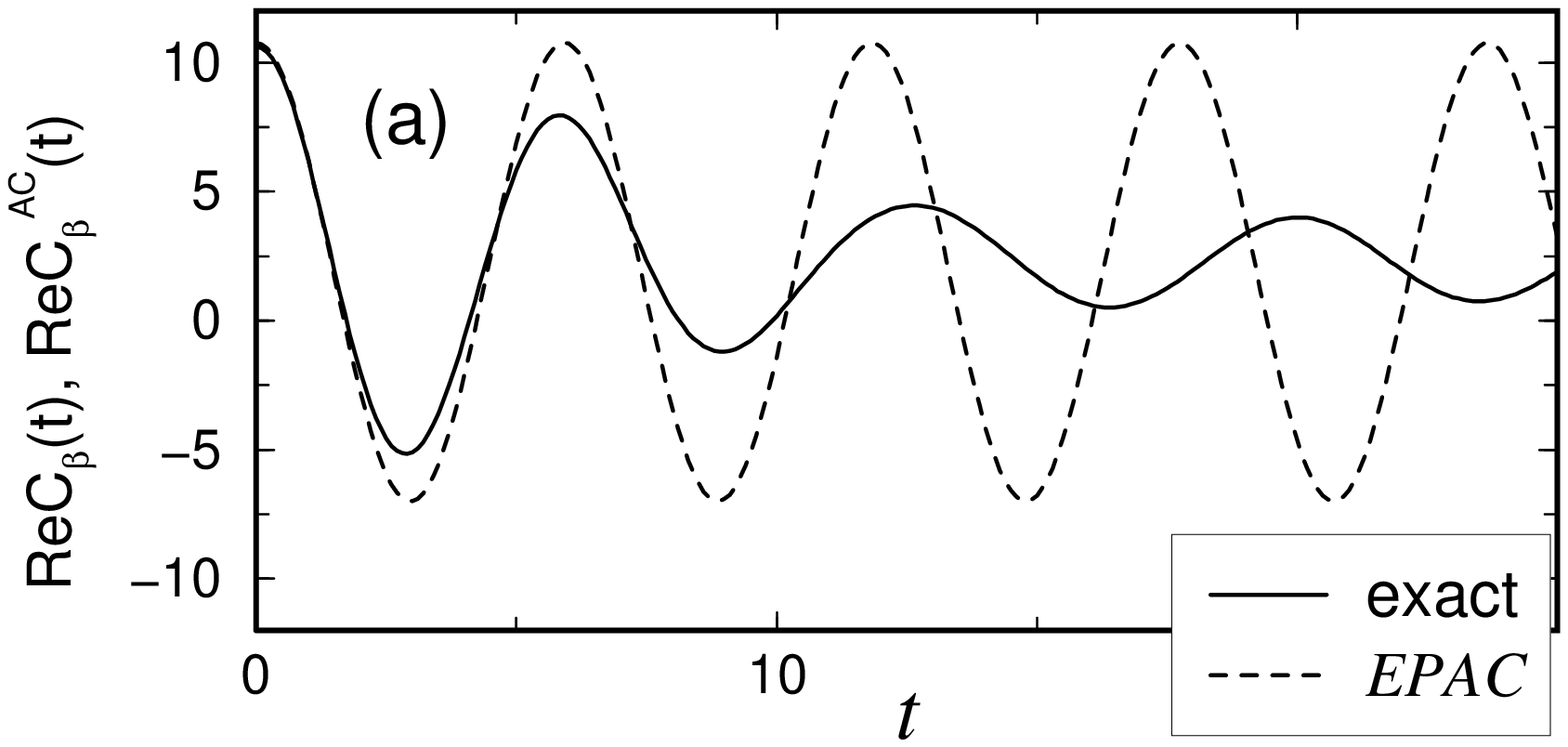}\\
\includegraphics[width=80mm]{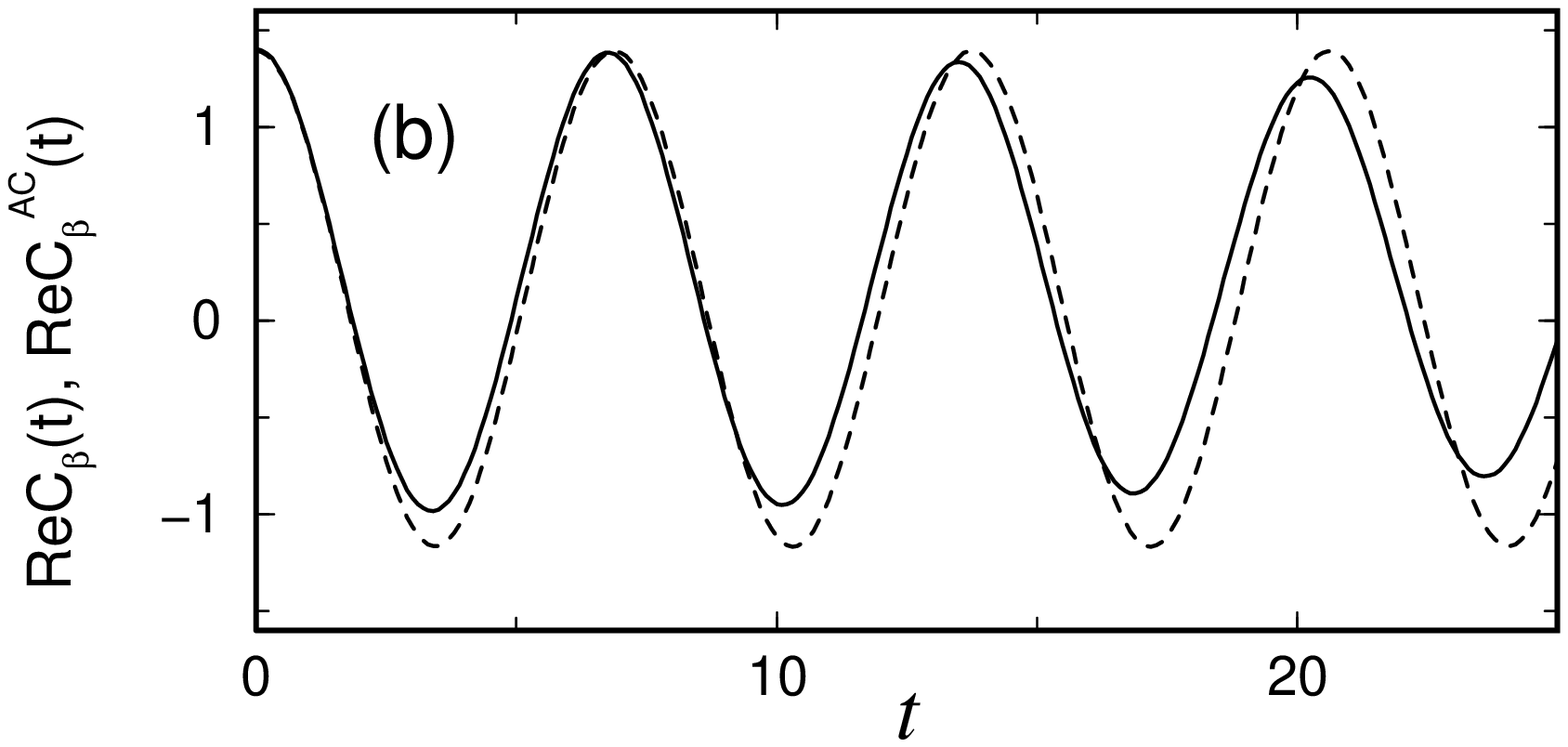}\\
\includegraphics[width=80mm]{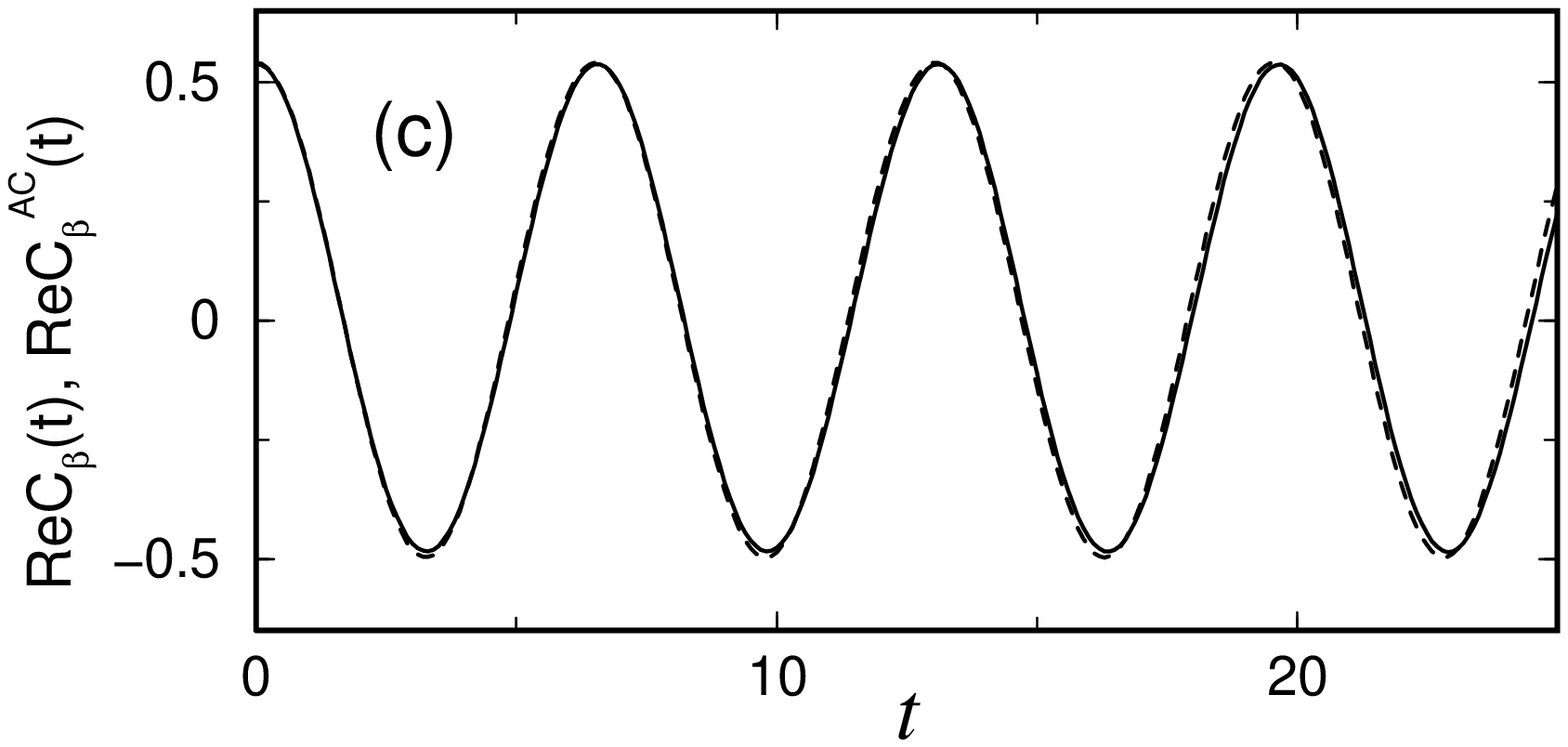}\\
\end{tabular}
\vspace{0mm}
\caption{
The real part of the exact quantum position autocorrelation function 
$C_{\beta}(t)$ 
and the real part of the EPAC position autocorrelation function 
$C_{\beta}^{\rm AC}(t)$:
(a) at $\beta=0.1$, (b) at $\beta=1$, and (c) at $\beta=10$. 
}
\label{fig:ACas}
\end{figure}

Finally we constructed 
the EPAC position autocorrelation functions $C_{\beta}^{\rm AC}(t)$ 
using the computed quantities $Q_{\rm min}$ and $\omega_{\beta}$.
Figure \ref{fig:ACas} shows
the real part of $C_{\beta}^{\rm AC}(t)$
together with the real part of 
the exact quantum position autocorrelation function,
\begin{eqnarray}
C_{\beta}(t)=\frac{1}{{\cal Z}_{\beta}}\sum_{n}\sum_{m}
e^{-\beta E_{n}}e^{-i(E_{m}-E_{n})t/\hbar}
\left|\langle m|\hat{q}|n\rangle\right|^{2},
\label{54}
\end{eqnarray}
at the inverse temperature $\beta=0.1$, 1, and 10.
We can see that each EPAC correlation function
reproduces well
the exact correlation function at $t=0$;
the static property $C_{\beta}(0)$
can be approximated very well by only two quantities 
$Q_{\rm min}$ and $\omega_{\beta}$,
\begin{eqnarray}
C_{\beta}(0)&\simeq& C_{\beta}^{\rm AC}(0)=
\frac{\hbar}{2m\omega_{\beta}}
\coth\frac{\beta\hbar\omega_{\beta}}{2}
+Q_{\rm min}^{2}
.\label{55}
\end{eqnarray}
This means that
the leading order derivative expansion 
employed in the EPAC \cite{hk}
is very good 
in the calculation of the static property $C_{\beta}(0)$. 
The value $C_{\beta}^{\rm AC}(0)$
can be improved by employing the higher order 
derivative expansion;
it should become exact in the infinite order expansion
if the expansion is a converging one.
On the other hand, as the time $t$ increases, 
each EPAC correlation function
deviates from the exact one.
Such deviation becomes larger at the higher temperature.
In particular, at $\beta=0.1$ [Fig. \ref{fig:ACas}(a)],  
the exact correlation function $C_{\beta}(t)$  
damps rapidly with time, whereas 
the EPAC correlation function $C_{\beta}^{\rm AC}(t)$  
shows stable oscillation. 
This damping behavior comes from the fact that 
the exact correlation function (\ref{54})
consists of the many oscillation modes with the frequencies 
$\omega_{m,n}=(E_{m}-E_{n})/\hbar$,
and oscillation modes with high energies contribute to Eq. (\ref{54})
more at higher temperature.  
On the other hand, the EPAC correlation function
consists of only one
oscillation mode with the frequency $\omega_{\beta}$
at any temperature, 
so that it always exhibits a single-mode oscillation.
As the temperature lowers [Fig. \ref{fig:ACas}(b) and \ref{fig:ACas}(c)], 
the EPAC correlation function $C_{\beta}^{\rm AC}(t)$
becomes closer to the exact $C_{\beta}(t)$.
This is because at lower temperature,
almost only one oscillation mode contributes to Eq. (\ref{54})
to make the exact $C_{\beta}(t)$ show a quasi-single-mode oscillation.
This property generally holds 
for a weakly anharmonic system \cite{cv,jv,nprg}
though it does not hold for strongly anharmonic systems
such as deep double well systems \cite{nprg}.    
\par
\par
The results shown in Fig. \ref{fig:ACas}
imply that
the leading order derivative expansion
employed in the EPAC method
becomes worse at longer time and at higher temperature. 
To reproduce the damping behavior of the exact $C_{\beta}(t)$ 
at high temperature,
we need the higher order derivative expansion of 
the effective action.
However, our calculation at lower temperature $\beta=10$
[Fig. \ref{fig:ACas}(c)]
corresponds to, for example, the dynamics of 
the intramolecular vibration of the HCl molecule
at temperature $T\sim 400 ~{\rm K}$.
Therefore our EPAC method 
even in the present form (\ref{22})
should work very well
for such chemical systems 
in the realistic temperature region.
\par
Finally we mention the  
applicability of the EPAC method to 
a nonpolynomial potential,
in particular, an asymptotically sublinear potential such as
$\lim_{|q|\to\infty}V(q)/|q|\leq 0$.
Most of the actual potentials appearing 
in chemical systems and condensed matter systems
have this type of potential shape,
for example, the Morse potential,
the Lenard-Jones potential, 
and other intramolecular and intermolecular potentials.
An important property of these potentials is that 
they represent the dissociation in the long distance limit. 
It is quite nontrivial to make 
the dissociation effect involved  
in the effective potential approach.
In fact, some difficulties have already been pointed out \cite{ct}. 
From the chemical interest,
this is a challenging subject to be examined.  

\section{CONCLUDING REMARKS}
\hspace*{\parindent}
In the present work we have focused on 
the EPAC analyses of one-dimensional asymmetric systems.
At first, 
for the asymmetric harmonic oscillator (\ref{41}), we have analytically 
shown that the exact real time position autocorrelation function 
is derived by means of the EPAC method.  
As for the asymmetric anharmonic oscillator (\ref{30}),
we could obtain the EPAC correlation function numerically.
In this case, at first
we have computed the standard effective potential $V_{\beta}(Q)$ 
by means of the PIMD technique and the numerical Legendre transformation.
Then we have obtained 
the position of the effective potential minimum $Q_{\rm min}$
and the effective frequency $\omega_{\beta}$
which are needed to construct   
the EPAC position autocorrelation function $C_{\beta}^{\rm AC}(t)$. 
It has been shown that 
the static properties of this asymmetric system,
$\langle\hat{q}\rangle_{\beta}$ and  
$C_{\beta}(0)$$(=\langle \hat{q}(0)\hat{q}(0)\rangle_{\beta})$,
are well reproduced by   
$Q_{\rm min}$ and $C_{\beta}^{\rm AC}(0)$, respectively.
As for a long time behavior,
we have found that
at lower temperature
the EPAC correlation function $C_{\beta}^{\rm AC}(t)$ agrees with 
the exact $C_{\beta}(t)$ very well,
whereas the EPAC result becomes worse 
as the temperature increases.
This is because the EPAC method approximates 
the multimode contribution to the exact real time correlation function
by the single-mode contribution
with the effective frequency $\omega_{\beta}$.  
In a chemical aspect, however, it is quite remarkable that
a good agreement of the EPAC correlation function with the exact one
is achieved for the realistic potential 
in the realistic temperature region.  
\par
We have also shown that a linear term in a classical potential 
is irrelevant to the 
effective potential calculation.
For a classical potential 
$V(q)=U(q)+fq$, 
the standard effective potential is given by
$V_{\beta}(Q)=U_{\beta}(Q)+fQ$.
We have proved this decoupling property analytically,
and have also checked it numerically 
by calculating $V_{\beta}(Q)$ and $U_{\beta}(Q)$ independently.

\par
In the present work, 
it has been shown that the EPAC method is practically applicable 
to simple systems with single degree of freedom.
More important applications in chemical physics are, however,
to the systems with many degrees of freedom.
For such applications, there are a couple of obstacles to be overcome.
The first one is   
the computation of the generating function $w_{\beta}(J)$. 
To obtain $w_{\beta}(J)$ using Eq. (\ref{7}), 
we need to precompute the effective classical potential 
$V_{\beta}^{c}(q_{c})$ itself. 
However, this  precomputation evidently demands more computational efforts    
as the system are more complicated.
For many-body systems, Eq.  (\ref{7a}) rather than Eq. (\ref{7})
will be useful for more efficient evaluation of $w_{\beta}(J)$, 
because the CMD sampling can be used to calculate the statistical-mechanical
average of $e^{\beta J q_{c}}$ appearing 
in the right-hand side of Eq. (\ref{7a}) \cite{comment2}.
That is, we can use the numerical techniques developed in
the CMD calculations for many-body systems \cite{cv,appli,appli2}. 
The second obstacle is the evaluation of 
the standard effective potential $V_{\beta}(Q)$
in many-body systems, i.e.,
the multidimensional Legendre transformation 
of $w_{\beta}(J)$ [Eq. (\ref{8})].  
This transformation will also require heavier computations     
for larger systems.
It is quite important to solve these numerical problems lying in 
the practical calculations for many-body systems.
We are presently developing the numerical schemes along this line. 

\begin{acknowledgments}
This work was supported by a fund for Research and Development 
for Applying Advanced Computational Science and Technology,
Japan Science and Technology Agency (ACT-JST).
\end{acknowledgments}



\begin{thebibliography}{99}
\bibitem{fh}
R. P. Feynman and A. R. Hibbs, 
{\it Quantum Mechanics and Path integrals}
(McGraw-Hill, New York, 1965);
R. P. Feynman,
{\it Statistical Mechanics}
(Addison-Wesley, New York, 1972).
\bibitem{bt}
B. J. Berne and D. Thirumalai,
Annu. Rev. Phys. Chem. {\bf 37}, 401 (1986),
and references cited therein.
\bibitem{ce}
D. M. Ceperley,
Rev. Mod. Phys. {\bf 67}, 279 (1995),
and references cited therein.
\bibitem{tb}
D. Thirumalai and B. J. Berne, 
J. Chem. Phys. {\bf 79}, 5029 (1983).
\bibitem{si}
N. Silver, J. E. Gubernatis, D. S. Sivia, and M. Jarrell,
Phys. Rev. Lett. {\bf 65}, 496 (1990);
R. N. Silver, D. S. Sivia, and J. E. Gubernatis,
Phys. Rev. B {\bf 41}, 2380 (1990);
J. E. Gubernatis, M. Jarrell, R. N. Silver, and D. S. Sivia,
{\it ibid}. {\bf 44}, 6011 (1991). 
\bibitem{gb}
E. Gallicchio and B. J. Berne, 
J. Chem. Phys. {\bf 101}, 9909 (1994); {\bf 105}, 7064 (1996);
E. Gallicchio, S. A. Egorov, and B. J. Berne,
{\it ibid}. {\bf 109}, 7745 (1998).
\bibitem{cv}
J. Cao and G. A. Voth, 
J. Chem. Phys. {\bf 99}, 10070 (1993);  
{\bf 100}, 5093 (1994); {\bf 100}, 5106 (1994); 
{\bf 101}, 6168 (1994); {\bf 101}, 6184 (1994);
G. A. Voth, 
Adv. Chem. Phys. {\bf XCIII}, 135 (1996).
\bibitem{hk}
A. Horikoshi and K. Kinugawa,
J. Chem. Phys. {\bf 119}, 4629 (2003).
\bibitem{fk}
R. P. Feynman and H. Kleinert, 
Phys. Rev. A {\bf 34}, 5080 (1986).  
\bibitem{gt}
R. Giachetti and V. Tognetti, 
Phys. Rev. Lett. {\bf 55}, 912 (1985); Phys. Rev. B {\bf 33}, 7647 (1986).
\bibitem{ep}
G. Jona-Lasinio,
Nuovo Cimento {\bf 34}, 1790 (1964).
\bibitem{riv}
R. J. Rivers,
{\it Path Integral Methods in Quantum Field Theory}
(Cambridge University Press, Cambridge, 1987).
\bibitem{ctvv}
A. Cuccoli, V. Tognetti, R. Vaia, and P. Verrucchi,
Phys. Rev. A {\bf 45}, 8418 (1992);
A. Cuccoli, V. Tognetti, P. Verrucchi, and R. Vaia,
Phys. Rev. B {\bf 46}, 11601 (1992); Phys. Rev. Lett. {\bf 77}, 3439 (1996);
A. Cuccoli, R. Giachetti, V. Tognetti, R. Vaia, and P. Verrucchi,
J. Phys.: Condens. Matter. {\bf 7}, 7891 (1995).
\bibitem{co}
S. Coleman, 
{\it Aspects of Symmetry} 
(Cambridge University Press, Cambridge, 1985).
\bibitem{jv}
S. Jang and G. A. Voth, 
J. Chem. Phys. {\bf 111}, 2357 (1999);  
{\bf 111}, 2371 (1999).
\bibitem{kb}
G. Krilov and B. J. Berne,
J. Chem. Phys. {\bf 111}, 9140 (1999);  
{\bf 111}, 9147 (1999).
\bibitem{ra}
R. Ramirez, T. Lopez-Ciudad, and J. C. Noya,
Phys. Rev. Lett. {\bf 81}, 3303 (1998);
R. Ramirez and T. Lopez-Ciudad,
{\it ibid}. {\bf 83}, 4456 (1999);
J. Chem. Phys. {\bf 111}, 3339 (1999). 
\bibitem{reich}
D. R. Reichman, P. -N. Roy, S. Jang, and G. A. Voth, 
J. Chem. Phys. {\bf 113}, 919 (2000).
\bibitem{appli}
J. Cao and G. J. Martyna,
J. Chem. Phys. {\bf 104}, 2028 (1996);
J. Cao, L. W. Ungar, and G. A. Voth,
{\it ibid}. {\bf 104}, 4189 (1996);
J. Lobaugh and G. A. Voth,
{\it ibid}. {\bf 104}, 2056 (1996); {\bf 106}, 2400 (1997);
M. Pavese and G. A. Voth,
Chem. Phys. Lett. {\bf 249}, 231 (1996);
A. Calhoun, M. Pavese, and G. A. Voth,
{\it ibid}. {\bf 262}, 415 (1996);
U. W. Schmitt and G. A. Voth,
J. Chem. Phys. {\bf 111}, 9361 (1999);
S. Jang, Y. Pak, and G. A. Voth, 
J. Phys. Chem. A {\bf 103}, 10289 (1999);
M. Pavese, S. Jang, and G. A. Voth, 
Parallel Comput. {\bf 26}, 1025 (2000);
U. W. Schmitt and G. A. Voth,
Chem. Phys. Lett. {\bf 329}, 36 (2000);
G. K. Schenter, B. C. Garett, and G. A. Voth,
J. Chem. Phys. {\bf 133}, 5171 (2000).
\bibitem{appli2}
K. Kinugawa, P. B. Moore, and M. L. Klein,
J. Chem. Phys. {\bf 106}, 1154 (1997); {\bf 109}, 610 (1998);  
K. Kinugawa,
Chem. Phys. Lett. {\bf 292}, 454 (1998);
S. Miura, S. Okazaki, and K. Kinugawa,
J. Chem. Phys. {\bf 110}, 4523 (1999);  
H. Saito, H. Nagao, K. Nishikawa, and K. Kinugawa,
{\it ibid}. {\bf 119}, 953 (2003).
Y. Yonetani and K. Kinugawa,
{\it ibid}. {\bf 119}, 9651 (2003).
\bibitem{proton} 
For example,
{\it Proton Transfer in Hydrogen-Bonded Systems},
edited by D. Bountis 
(Plenum, New York, 1992);
{\it Electron and Proton Transfer in Chemistry and Biology},
edited by A. M\"uller, H. Ratajczak, W. Junge, and E. Diemann
(Elsevier, Amsterdam, 1992);
{\it Ultrafast Hydrogen Bonding Dynamics and 
Proton Transfer Processes in the Condensed phase},
edited by T. Elsaesser and H. J. Bakker
(Kluwer Academic, Dordrecht, 2002).
\bibitem{gold}
N. Goldenfeld, 
{\it Lectures on Phase Transitions and the Renormalization Group} 
(Addison-Wesley, Reading, MA, 1992). 
\bibitem{asym}
S. Seide and C. Wetterich,
Nucl. Phys. B {\bf 562},524 (1999); 
A. Strumia and N. Tetradis,
{\it ibid}. {\bf 542}, 719 (1999).
\bibitem{kl}
H. Kleinert, 
{\it Path Integrals in Quantum Mechanics Statistics and Polymer Physics}
(World Scientific, Singapore, 1995).
\bibitem{fukuda}
R. Fukuda and E. Kyriakopoulos, 
Nucl. Phys. B {\bf 85}, 354 (1975);
R. Fukuda,
Prog. Theor. Phys. {\bf 56}, 258 (1976).
\bibitem{hs}
U. M. Heller and N. Seiberg,
Phys. Rev. D {\bf 27}, 2980 (1983).
\bibitem{owy}
L. O'Raifeartaigh, A. Wipf, and H. Yoneyama,
Nucl. Phys. B {\bf 271}, 653 (1986).
\bibitem{bell}
M. Le Bellac,
{\it Thermal Field Theory}
(Cambridge University Press, Cambridge, 1996).
\bibitem{comment0}
The CMD method is also exact for harmonic systems because 
the canonical (Kubo-transformed) 
position autocorrelation function
$C_{\beta}^{\rm can}(t)=(1/\beta)\int_{0}^{\beta}d\lambda
~\langle \hat{q}(t-i\hbar\lambda)\hat{q}(0)\rangle_{\beta}
=(1/\beta m\omega^{2})\cos\omega t$
can be exactly reproduced by using the classical dynamics on the 
effective classical potential surface (\ref{25}). 
\bibitem{comment1}
To be more precise, 
the approximated form of the Morse potential (\ref{31}) is given by
$V(q)=\frac{1}{2}~\!q^{2}+
\frac{1}{10}~\!q^{3}+
\frac{7}{600}~\!q^{4}$.
\bibitem{herz}
G. Herzberg,
{\it Molecular Spectra and Molecular Structure. 
I. Spectra of Diatomic Molecules}, 2nd ed.
(Van Nostrand, New York, 1950).
\bibitem{nprg}
K.-I. Aoki, A. Horikoshi, M. Taniguchi, and H. Terao,
Prog. Theor. Phys. {\bf 108}, 571 (2002).
\bibitem{comment2}
Using not Eq. (\ref{7}) but Eq. (\ref{7a}), however, 
we can skip the precomputation of the 
effective classical potential itself. 
Equation (\ref{7a}) allows us to
directly evaluate the generating function 
$w_{\beta}(J)$ 
as a statistical-mechanical average of $e^{\beta J q_{c}}$
by means of the canonical CMD sampling 
(Refs. \onlinecite{cv}, \onlinecite{appli}, and \onlinecite{appli2}) 
of the centroid variables ($q_{c}$, $p_{c}$).  
\bibitem{ct}
T. L. Curtright and C. B. Thorn, 
J. Math. Phys. {\bf 25}, 541 (1984).  
 
\end{thebibliography}
\end{document}